\documentclass[aps,prl,twocolumn,groupedaddress,showpacs]{revtex4}

\usepackage{graphicx}
\usepackage{amsmath}

\begin{document}

\title{Diffraction of a Bose-Einstein Condensate in the Time Domain}

\author{Yves Colombe}
\altaffiliation[Present address: ]{Sektion Physik der Ludwig-Maximilians-Universit\"at, Schellingstrasse 4, D-80799 M\"unchen, Germany.}
\author{Brigitte Mercier}
\altaffiliation[Present address: ]{Laboratoire d'Optique Appliqu\'ee, Centre de l'Yvette, F-91761 Palaiseau Cedex, France.}
\author{H\'el\`ene Perrin}
\author{Vincent Lorent}
\email{lorent@galilee.univ-paris13.fr}
\affiliation{Laboratoire de Physique des Lasers, UMR 7538 du CNRS, Institut Galil\'ee, Universit\'e Paris-Nord, Avenue J.-B. Cl\'ement, F-93430 Villetaneuse, France}

\date{\today}

\begin{abstract} We have observed the diffraction of a Bose-Einstein condensate of rubidium atoms on a vibrating mirror potential. The matter wave packet bounces back at normal incidence on a blue-detuned evanescent light field after a $3.6\,$mm free fall. The mirror vibrates at a frequency of $500\,$kHz with an amplitude of $3.0\,$nm. The atomic carrier and sidebands are directly imaged during their ballistic expansion. The locations and the relative weights of the diffracted atomic wave packets are in very good agreement with the theoretical prediction of Carsten Henkel \textit{et al.} \cite{Henkel1994}.
\end{abstract}


\maketitle

The manipulation of ultracold atomic matter waves with optical or magnetic fields close to surfaces is extensively explored in the context of fabricating integrated atom optics devices. The use of the Zeeman interaction due to the magnetic field of microfabricated current carrying wires is currently the most attractive approach \cite{Reicheletal2001,Folmanetal2002}. The main advantages of this method are the modularity and steadiness of the microchip fabrication. Nevertheless, some drawbacks of this technique exist since one experiences losses of atoms in magnetic traps at close distances to conducting surfaces \cite{Jonesetal2004,Linetal2004,Henkel1999} due to Johnson noise induced spin flips. This loss mechanism is absent in the vicinity of dielectric surfaces, which can be used as substrates for dipole traps based on optical near fields. In 1991 Ovchinnikov \textit{et al.} \cite{Ovchinnikov1991} made the seminal proposal of using the difference in the decay lengths of the evanescent fields created by total internal reflections of blue- and red-detuned light beams on a planar dielectric surface to create a trapping dipole potential above the surface. The group of R. Grimm demonstrated this trapping in 2002 \cite{Hammes2003}. The proposals of Barnett \textit{et al.} \cite{Barnettetal2000} and Burke \textit{et al.} \cite{Burkeetal2002} enlarge the optical near field trapping geometry to a richer variety of patterns: the basic idea is to take benefit of light injected inside integrated optical structures to design evanescent field traps and guides. Having a similar compactness and versatility as the optical waveguides supporting the evanescent waves, these dipole traps and guides offer an interesting alternative to the current carrying wires on a chip technique.

In this letter we address the action of the evanescent outer part of a light mode propagating in a planar optical waveguide. The experiment performed is similar to the one realized with cold atoms by the group of Jean Dalibard in 1995 \cite{Steane1995}. The difference mostly consists in the initial longitudinal coherence of our atomic source.
In our experiment, an atomic Bose-Einstein condensate is reflected after a free fall by the evanescent part of a blue-detuned guided optical mode, and is observed in its ballistic expansion after the bounce. The evanescent mirror is made to vibrate, which modulates the phase of the reflected wave function and diffracts the atoms in several sidebands. Atoms bouncing on this potential are also dramatically scattered, which is due to the corrugated structure of the planar waveguide on a nanometer scale. A particular study of the elastic scattering of the atomic wave by the static rough mirror potential is presented elsewhere \cite{Perrin2005}. As discussed below, the diffraction in the time domain and the elastic scattering are independent phenomena and we focus here on the study of the first effect. 
This diffraction of an atomic BEC by a modulated mirror has also similarities with experiments on atomic diffraction performed by a pulsed optical standing-wave \cite{Ovchinnikovetal1999, Kelleretal1999}.
The experimental set-up is described in \cite{Colombe2003}. It is based on a double MOT system. From the UHV MOT $5.10^8$ $^{87}$Rb atoms are transferred into a QUIC (Quadrupole and Ioffe Configuration) \cite{Esslinger1998} magnetic trap. An almost pure condensate of $2.10^5$ atoms is obtained by radio-frequency evaporative cooling inside the QUIC trap. Below the trapped atoms stands an optical waveguide made of a $360\,$nm thick layer of TiO${}_2$ ($n_\text{guide} = 2.3$) on the top of a $400\,$nm SiO${}_2$ layer ($n_\text{gap} = 1.46$). This low index gap layer is on the top surface of a high index prism made of a Schott LaSFN15 glass ($n_\text{prism}=1.86$). The TE2 mode of the waveguide is excited through evanescent coupling of a $P_0 = 50\,$mW diode laser beam detuned by $\delta_0 = 2.1\,$GHz on the blue side of the $\mathrm{D}_2$ $\mathrm{5S}_{1/2},\:F=2$ $\mathrm{\rightarrow 5P}_{3/2}$ transition. The field decay length of the TE2 guided mode in the vacuum is $\kappa^{-1} = 93\,$nm. The number of spontaneous photon per atom during the bounce is about $0.1$ in the situation where the atoms penetrate in the evanescent field with a falling height of $3.6\,$mm. The vibration of the evanescent mirror is obtained by a sinusoidal modulation of the diode current. The resulting modulation depth of the a.c. Stark shift potential $U \propto P/\delta$ is $\varepsilon = | \varepsilon_P + \varepsilon_\delta |$, where $\varepsilon_P = \Delta P/P_0$ and $\varepsilon_\delta = -\Delta \delta/\delta_0$. The power modulation depth $\varepsilon_{P}$ is measured directly with a photodiode. The detuning modulation depth $\varepsilon_{\delta}$ is measured by calibrating the frequency shift versus the diode current using the atomic frequency reference in a saturated absorption experiment. The modulation depth $\varepsilon_{P}$ is $25$ times less than $\varepsilon_{\delta}$: the modulation of the reflecting potential is thus essentially due to the modulation of the detuning.
\begin{table}
\caption{\label{table:Parameters}Parameters for the (a), (b), and (c) experiments: frequency modulation $\Omega/2\pi$, optical frequency detuning $\delta_{0}/2\pi$, detuning modulation $\Delta\delta/2\pi$, modulation depth $\varepsilon$, fall height $z_{0}$ and time of flight $\Delta t_\text{fall}+\Delta t_\text{bounce}$.}

\vspace{2mm}
\begin{tabular}{c c c c}
\hline\hline
Experiment                        & (a)         & (b)       & (c)         \\
\hline
$\Omega/2\pi\;(\text{kHz})$       & $500$       & $500$     & $500$       \\
$\delta_{0}/2\pi\;(\text{GHz})$   & $+2.1$      & $+2.1$    & $+1.9$      \\
$\Delta\delta/2\pi\;(\text{MHz})$ & $130$       & $163$     & $163$       \\
$\varepsilon$                     & $6.2\,\%$   & $7.8\,\%$ & $8.6\,\%$   \\
$z_{0}\;(\text{mm})$              & $3.6$       & $3.6$     & $2.05$      \\
$\Delta t_\text{fall}+\Delta t_\text{bounce}\;(\text{ms})\;$ & $27+27\;\;$ & $27+27\;$ & $20.5+19.5$ \\
\hline
\end{tabular}
\end{table}

\begin{figure*}[ht]
\includegraphics[width=18 cm]{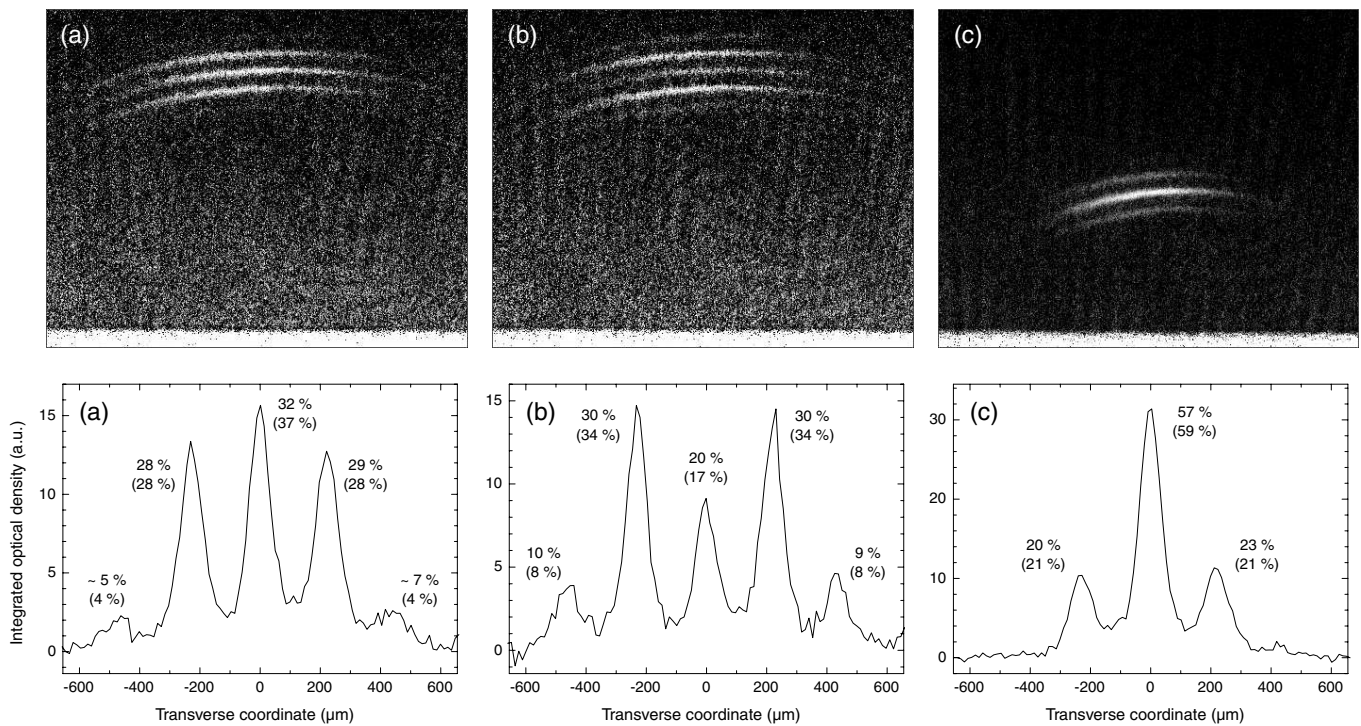}
\caption{\label{fig:Diffraction}Diffraction in the time domain of a $^{87}\text{Rb}$ BEC on a vibrating repulsive evanescent wave. The experimental parameters for (a), (b), and (c) are summarized in table~\ref{table:Parameters}. The circular shape of each diffracted order is due to elastic scattering of the atoms on the rough mirror potential (see text). Each absorption image is the result of a single experimental run. The camera field is $5.5\,$mm (hor.) $\times$ $4.4\,$mm (vert.) and includes the top of the prism. The slight tilt of the atomic clouds is due to a small horizontal initial velocity of $30\,$mm/s. The values on the optical density profiles are the relative weights of the diffracted orders; expected weights are in parentheses.\vspace{-2mm}}
\end{figure*}
Three kinds of BEC diffraction experiments have been performed. They differ by the falling heights of the atomic cloud and by the modulation depths of mirror potential. The parameters are summarized in table~\ref{table:Parameters}. In the (a) and (b) experiments the atomic condensate is released from the QUIC trap centered $3.6\,$mm above the dielectric surface. In the (c) experiment the condensate is first magnetically transported to a height of $2.05\,$mm above the surface before being released from the trap \cite{Colombe2004}. The corresponding free fall times down to the dielectric surface are $\Delta t_\text{fall}=27\,$ms in cases (a,b) and $20.5\,$ms in case (c). In addition to the incident vertical velocity, a small horizontal velocity ($30\,$mm/s for all experiments) results in a slight tilt of the clouds after reflection. The modulation depths $\varepsilon$ are $6.2\,\%$ in (a), $7.8\,\%$ in (b), and $8.6\,\%$ in (c). These modulation depths correspond to vibrating amplitudes of the mirror $z_\text{M}$ of $2.9\,$nm, $3.6\,$nm, and $4.0\,$nm, respectively. The frequency of the modulation is kept the same, $\Omega = 2\pi \times 500\,$kHz, throughout these experiments. The diffracted wavepackets are detected by absorption imaging with a horizontal laser beam (Fig.~\ref{fig:Diffraction}, upper part). In order to accurately measure the atomic populations in the different elastically scattered diffraction orders, the optical density is integrated along circles of growing radii. The relative weights are measured on the resulting profile (Fig.~\ref{fig:Diffraction}, lower part) \cite{Circles}.

\begin{table}[t]
\caption{\label{table:Positions}Expected and measured positions of the sidebands relative to the carrier, in $\mu$m, for experiments (a), (b), (c).}

\vspace{2mm}
\begin{tabular}{c c c c c c}
\hline\hline
Diffraction orders    & $-2$   & $-1$   & $0$     & $+1$   & $+2$ \\
\hline
Expected: (a) and (b) & $-479$ & $-235$ & $\;0\;$ & $+228$ & $+449$ \\
Measured: (a)         & $-470$ & $-226$ & $0$     & $+221$ & $+433$ \\
Measured: (b)         & $-460$ & $-231$ & $0$     & $+219$ & $+433$ \\
\hline
Expected: (c) 				&				 & $-228$ & $0$     & $+216$  \\
Measured: (c)					&				 & $-227$ & $0$     & $+218$  \\
\hline
\end{tabular}
\end{table}
The diffracted populations are clearly resolved with a time of flight $\Delta t_\text{bounce}=27\,$ms (a,b) or $19.5\,$ms (c) after the reflection on the evanescent mirror.
The measured distances between the diffracted orders are reported in table~\ref{table:Positions}. A remarkable feature of this experiment lies in the direct visualization of the sidebands. The wavenumber separations are transferred into wave packet separations that allow a direct and accurate measurement of the energy intervals and relative weights of the sidebands. The scattering of the matter wave by the mirror roughness and the diffraction of the same matter wave by the mirror vibration are different in nature and their effects are indeed clearly separated on the absorption images. The first effect is an elastic momentum exchange which spreads the reflected atoms over an elastic scattering sphere. The second is a transfer of energy, giving birth to sidebands. Given our particular condition of initial kinetic energy for say, $3.6\,$mm free fall, the velocity difference along $z$ between the carrier and the first sidebands is $\Delta v \simeq \pm 1.5\, v_\text{rec}$ ($v_\text{rec}$ is the photon recoil velocity) with a modulation frequency of $500\,$kHz. The momentum scattering due to diffuse reflection affects essentially the horizontal velocity ($\sigma _v = 6.6\, v_\text{rec}$ along $x$ \cite{Perrin2005}). Hence it does not prevent the observation of resolved sidebands along the vertical direction. This leads us to interpret our diffraction experiments with a one-coordinate model, namely an incident plane wave at normal incidence with a perfect mirror.

Let us first consider the reflection of the plane atomic wave function on a vibrating infinite repulsive potential $U \bigl( z<z_\text{m}(t) \bigr) = +\infty\,,\;U \bigl( z\geq z_\text{m}(t) \bigr) = 0$ \cite{WavePacket}. We assume that the velocity of the mirror coordinate $z_\text{m}$ is always much less than the atomic group and phase velocities, so that the incident wave function can be written as $\psi _i(z\geq z_\text{m},t) = \exp [i(-kz-\omega t)]$. The incident and reflected waves fulfill the boundary condition $\psi_\text{i}(z_\text{m}(t),t)+\psi_\text{r}(z_\text{m}(t),t)=0$ at any time. In the case of a harmonically vibrating mirror $z_\text{m}(t)=z_\text{M}\sin(\Omega t)$, the reflected wave function may be expanded as a sum of a carrier and diffracted sidebands $\psi _r \bigl( z\geq z_\text{m}(t),t \bigr) = \sum\limits_{n = -\infty}^{+\infty} J_n(2kz_\text{M})\,\exp i[ k_n z -( \omega + n\Omega )\,t + \pi ]$, where the sideband amplitudes are Bessel functions of the first kind and the atomic phase modulation amplitude is $2kz_\text{M}$. The energy separation between sidebands is $\hbar \Omega$ and the corresponding wavenumbers are $k_{n}\simeq k+n\frac{\Omega M}{\hbar k}$ ($M$ is the mass of the atom) as long as the energy transfer is much less than the incident kinetic energy.

In our situation the reflecting potential is an exponential whose amplitude is harmonically modulated $U(z,t)=U_{0}[1+\varepsilon\sin(\Omega t)]\exp(-2\kappa z)$. As the potential is exponential, the amplitude modulation is equivalent to an overall translation; in the case of a weak modulation depth $\varepsilon\ll 1$, this translation is also harmonic: $U(z,t)=U_{0}\exp\bigl[-2\kappa\bigl(z-z_\text{M}\sin(\Omega t)\bigr)\bigr]$ with $z_\text{M}=\frac{\varepsilon}{2\kappa}$. The main difference between the infinitely steep and the evanescent potentials lies in the continuous variation of the incident matter wave momentum inside the potential in the last case. Henkel \textit{et al.} have calculated the atomic phase modulation imprinted by the vibrating mirror in a semiclassical model \cite{Henkel1994}. It is assumed that the incident atomic de Broglie wavelength is much smaller than $2\pi\kappa^{-1}=585\,$nm, and that the classical atomic trajectory is not much affected by the vibration of the potential. In our experiment the atomic BEC cloud is released $3.6\,$mm or $2.05\,$mm above the dielectric substrate. When the atoms hit the evanescent mirror, the de Broglie wavelength $\lambda _\text{dB}$ is respectively $17\,$nm or $23\,$nm. These values are indeed much smaller than $2\pi\kappa^{-1}$. Furthermore, our modulation depth is at maximum $ \varepsilon=8.6\,\%$ and ensures that the vibration barely perturbs the classical atom trajectory. Under these conditions the semi-classical approach proposed by Henkel \textit{et al.} is valid.

The predicted diffraction weights are
\begin{equation}
\mathcal{P}(n) = \bigl|J_n \bigl[ 2kz_\text{M} \beta (Q) \bigr]\bigr|^2 \label{eq:Weights}
\end{equation}
where $\beta(x) = \frac{\tfrac{\pi}{2}x}{\sinh(\tfrac{\pi}{2}x)}$ and $Q=\frac{\Omega M}{\hbar k}/ \kappa$. $Q$ is the ratio of the wavenumber interval between successive sidebands and the exponential decay factor of the evanescent wave. The reduction factor $\beta(Q)$ falls exponentially for $Q>1$, so that the maximum momentum transfer is in the order of $\hbar\kappa$ as expected from the Heisenberg uncertainty relation. The values in parentheses in Fig.~\ref{fig:Diffraction} are the weights $\mathcal{P}(n)$ calculated by the formula~\eqref{eq:Weights} where the experimental values serve as inputs. The agreement between the calculated and the observed weights is within $10\,\%$ accuracy in the worst case.
\begin{figure}[ht]
\begin{center}\includegraphics[width=0.95\columnwidth]{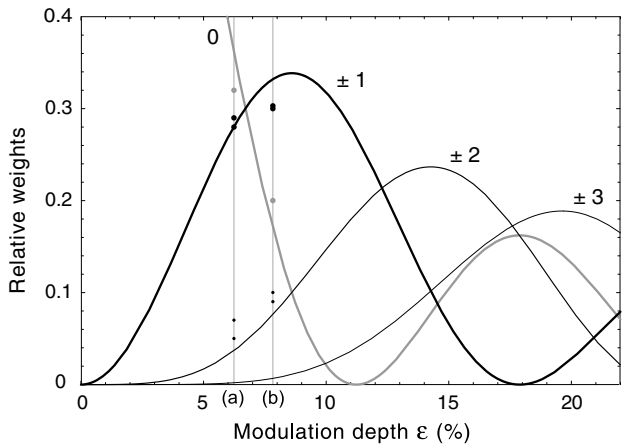}
\end{center}
\caption{\label{fig:Weights}Relative weight distribution over the carrier and the first six sidebands of an atomic wave packet reflected off the vibrating evanescent mirror for a $500\,$kHz modulation, as a function of the modulation depth $\varepsilon$. The curves are the weights $\mathcal{P}(n)$ \eqref{eq:Weights} given by the semiclassical model of Henkel \textit{et al.} \cite{Henkel1994} for $^{87}$Rb atoms released $3.6\,$mm above the dielectric surface. The values plotted at modulations depths $6.2\%$ and $7.8\%$ are those of the (a) and (b) experiments, respectively (see table~\ref{table:Parameters} and Fig.~\ref{fig:Diffraction}).}
\end{figure}
The figure~\ref{fig:Weights} illustrates the expected relative weights of the diffracted orders as a function of the modulation depth $\varepsilon$ for atoms falling a height of $3.6\,$mm and a mirror modulation frequency of $500\,$kHz. It clearly shows that small modulation depths are the better choice to combine high diffraction efficiency on a few sidebands.

In summary, this experiment demonstrates the diffraction of an atomic matter wave by a vibrating rough mirror potential. Despite the elastic diffusion of the atoms, the signal of diffraction is still clear-cut because of the monochromaticity of the atomic source. However, it would be misleading to associate the quantum nature of the atomic diffraction to the Bose-Einstein phase coherence. The dynamics of the bouncing is not even determined by the density term of the Gross-Pitaevskii equation as it is in the Hannover experiment \cite{Bongs1999}: the expanding BEC evolves like a free non interacting particle gas already $2\,$ms after being released from the magnetic trap. The linear Schr\"odinger equation gives correctly the dynamics of individual atoms, independently of a relative phase between their wave functions. Such a vibrating mirror can be used as a phase modulator in conventional atom optics: it has been implemented in a longitudinal interferometer with three consecutive bounces \cite{Szriftgiser1996}, the temporal equivalent of three grating interferometers \cite{Keith1991,Rasel1995}. In these devices, the atomic sources are considered as white light sources and great care is taken to have identical path lengths. Crossing the bridge to non-symmetric path interferometry becomes realistic when the atomic wave comes out of a BEC. Under our experimental conditions, a 2 path interferometer seems realistic under the following conditions: a first separation of order $+1$ and $-1$ followed by $N$ and $N+1$ bounces respectively, the final recombination being ensured by a last modulation, would lead to a very asymmetric interferometer. With our experimental parameters, $N=7$ appears to be possible. It will require, however, the use of a super-polished substrate as a mirror and possibly a lateral confinement of the atomic wave: up to $10$ bounces have been observed with a conventional MOT atomic source at a $3\,$mm drop height above a curved mirror \cite{Aminoff1993}, and a guiding of the matter wave without perturbing the motion perpendicular to the mirror surface can be obtained with magnetic confinement \cite{Dekkeretal2000}.

\begin{acknowledgments}
We thank B.~M.~Garraway for a careful reading of the manuscript. We acknowledge support from the FNS through the ACI-photonique program, the UE through the RTN FASTNet program under contract No. HPRN-CT-2002-00304 and the Conseil R\'egional d'Ile de France.
\end{acknowledgments}

\bibliography{Colombe}

\begin{thebibliography}{25}
\bibitem[{Cir()}]{Circles}
\bibinfo{note}{For each experiment, the center of the integration circles
  corresponds to the point of contact of the atoms with the evanescent wave,
  translated vertically by the distance $\frac{1}{2}g\Delta t_\text{bounce}^2$
  covered by the free fall referential that is at rest at the instant of the
  reflection. The wave packets are adequately fitted with Gaussian functions,
  which allow a measure of their relative weights and positions.}
\bibitem[{Wav()}]{WavePacket}
\bibinfo{note}{The spatial width of the incident wave packet is still large
  enough to ensure the validity of the plane wave model.}
\end{thebibliography}

\end{document}